\documentclass[%
 reprint,
 amsmath,amssymb,
 aps,
pre,
]{revtex4-1}

\usepackage{graphicx}
\usepackage{dcolumn}
\usepackage{bm}
\usepackage{placeins}

\begin{document}

\preprint{APS/123-QED}

\title{Non-equilibrium odds for the emergence of life}

\author{Elan Stopnitzky}
\email{elanstop@hawaii.edu}

 \affiliation{%
 Department of Physics, University of Hawaii at Manoa
}%

\author{Susanne Still}%
 \email{sstill@hawaii.edu}
\affiliation{Department of Information and Computer Sciences, University of Hawaii at Manoa\\
}%


%


\date{\today}

\begin{abstract}
Large and complex molecules are building blocks for life. We compute probabilities for their formation from an average non-equilibrium model. 
As the distance from thermodynamic equilibrium is increased in this model, so too are the chances for forming molecules that would be prohibitively rare in thermodynamic equilibrium. This effect is explored in two settings: the synthesis of heavy amino acids, and their polymerization into peptides. In the extreme non-equilibrium limit, concentrations of the heaviest amino acids can be boosted by a factor of 10,000. 
Concentrations of the longest peptide chains can be increased by hundreds of orders of magnitude. 
Since all details of the non-equilibrium driving are averaged out, these findings indicate that, independent of the details of the driving, the mere fact that pre-biotic environments were not in thermodynamic equilibrium may help cross the barriers to the formation of life.  
\end{abstract}

\maketitle


\section{\label{sec:level1}Introduction}

Biology requires the coordination of many complex molecules to store and copy genetic information, harness energy from the environment, and maintain homeostasis. 
The emergence of life thus hinges upon the likelihood of such molecules originating from an abiotic environment. 
At first glance, statistical mechanics seems to pose a barrier to this program: the high molecular mass and structural specificity of many biomolecules severely limit the likelihood of their spontaneous formation in thermodynamic equilibrium and thus make the spontaneous emergence of life implausible \cite{enzyme,typewriter,mineral,critparams,review}. 

The severity of this problem, which appears under rather general considerations, has motivated researchers to search for special environments, either extant or belonging to the early Earth, which would be ideally suited to produce the needed molecules in significant quantities. 
Examples of such environments include hydrothermal vent systems \cite{hydrotherm,hydrorev,alkavent,amend1998energetics}, and the surfaces of minerals \cite{mineral}. This approach is impeded in part by uncertainty about the chemistry of early Earth \cite{Bada, presynth,review}. Moreover, the set of organisms from which we derive our understanding of biochemistry is at least partly the result of historical chance. 
Even a very convincing account would not suffice to rule out the possibility of life forming under different conditions. This becomes a serious problem if life is a more general phenomenon than the available examples suggest.

To form, many biomolecules require free energy, and non-equilibrium driving of some kind is imperative for synthesis to occur \cite{enzyme,mineral,critparams,review,eric,geodiseq}. Some proposed sources of this driving in the pre-biotic Earth are radiation \cite{geodiseq,eric}, temperature and ion gradients \cite{gradient,geodiseq,eric}, concentration fluxes \cite{copoly,rob}, and electrical discharge \cite{Miller245}.  




Rather than looking for specific conditions that might have created life, we want to ask a simple, more general question: how much would non-equilibrium conditions typically change the chances of forming the complex molecules that life relies on? To that end, we consider the {\em average} non-equilibrium distribution \cite{Crooks}, which allows us to compute odds for the formation of biologically important molecules without having to make any specific assumptions. This calculation predicts that the odds can be increased significantly, depending on how far from equilibrium conditions are assumed to have been in pre-biotic environments. We illuminate this effect using simple models for the spontaneous synthesis of heavy amino acids (Sec. \ref{amino}), and for their polymerization into peptides (Sec. \ref{peptide}).

The role of non-equilibrium driving elucidated in the literature \cite{enzyme,mineral,critparams,review,eric,gradient,copoly,geodiseq,drivetolife} can thus be seen as part of a much more general phenomenon, whereby system states that are comparatively rare in equilibrium typically become more probable further away from equilibrium. This effect can augment the probabilities of forming rare molecules by many orders of magnitude, and therefore may help to bridge some of the most serious gaps in our understanding of the origin of life.

\section{The Non-Equilibrium Model} \label{noneq}
\label{hyper}

Estimating the average non-equilibrium distribution is by no means an obvious endeavor. It requires theoretical guidance, and, along the way, certain assumptions. In this paper, we follow a framework proposed by Crooks \cite{Crooks} and explained in the Methods section, which leads us to the following expression for the average non-equilibrium distribution:
\begin{equation} \label{eq:noneq}
\langle \theta \rangle \sim \int \theta e^{ -\lambda D (\theta \parallel \rho)} d \theta ~.
\end{equation}
The relative entropy between distribution, $\theta$, and the corresponding equilibrium distribution, $\rho$,
\begin{equation}
 D ( \theta || \rho) = \sum_i \theta_i \ln \left[ \frac{\theta_i}{\rho_i} \right],
\end{equation}

\noindent measures the additional free energy available in a non-equilibrium distribution \cite{shaw1984dripping,takara},
and it measures the inefficiency encountered when the canonical (equilibrium) distribution $\rho$ is used as a model for $\theta$ \cite{kullback1959, elements}. 

The parameter $\lambda$ reflects the distance away from thermodynamic equilibrium. Of course we do not know how far pre-biotic Earth was out of equilibrium, and thus we can not determine the parameter $\lambda$. We can, however, gain valuable insights by studying probabilities derived from the average non-equilibrium distribution, $\langle \theta \rangle$, as a function of $\lambda$.
 In the limit $\lambda \rightarrow \infty$, the average non-equilibrium distribution does not differ from the equilibrium distribution: $\langle \theta \rangle =\rho$. In the other limit, extremely far from equilibrium, $\lambda =0$, and the average non-equilibrium distribution becomes flat: $\langle \theta \rangle = const.$

For finite $\lambda$ values that are not too large, the distribution $\langle \theta \rangle$ is in general flatter than its equilibrium counterpart, thereby augmenting the probabilities of states that would otherwise be rare \cite{Crooks}. This important effect should have profound implications for our understanding of the origin and evolution of life, as a myriad of biological processes seem to rely on the chance occurrence of fantastically improbable events. In the following sections we calculate the average non-equilibrium distribution (Eq. \ref{eq:noneq}) for two biologically relevant model systems, and show how the odds of forming large and complex molecules are boosted for non-equilibrium systems.

 \section{Amino Acid Abundances and Functional Proteins} \label{amino}
The possibility of pre-biotic synthesis of amino acids was established in the landmark experiment by Miller and Urey \cite{Miller245}. They have since been detected in meteors \cite{pizza}, and produced in other experiments seeking to model the conditions of the early Earth \cite{Bada,McCollom}. However, the abundances with which the amino acids appear in abiotic settings do not match their biotic abundances \cite{Adami}. In particular, functional proteins tend to employ the various amino acids in roughly equal proportions \cite{Adami, adami-information}, whereas in abiotic sources there is an exponential suppression in the abundances of the larger amino acids, and none heavier than threonine have yet been found \cite{aminotherm}. The apparent inability of the environment to produce heavier amino acids in sufficient quantities has been identified by several authors as a barrier to the emergence of life \cite{adami-information,aminotherm,review}. 


The difficulty of synthesizing the heavier amino acids in a pre-biotic setting is usually ascribed to them having a larger Gibbs free energy of formation, $\Delta G$ {\cite{aminotherm}. The free energies of formation of the amino acids were calculated in \cite{amend1998energetics}, assuming synthesis from $CO_2$, $NH_4^{+}$, and $H_2$ in surface seawater at a temperature of $18^{\circ}C$. The concentrations of amino acids relative to glycine, taken from 9 different data sets, were fit using an exponential function \cite{aminotherm}:
\begin{equation} \label{eq:empi}
 C_{\rm rel}=15.8 *\mbox{exp} \left[-\Delta G/31.3 \right]~.
\end{equation}
We rescale these values so that they may be interpreted as probabilities (i.e. fraction of material in the solution):


\begin{equation} \label{eq:pro}
 P(x)=\frac{C_{\rm rel}(x)}{\sum_{i=1}^N C_{\rm rel}(i)}~,
\end{equation}

\noindent where $C_{\rm rel}(x)$ is the relative concentration of amino acid $x$, and the index $i=1, \dots, N$ runs over all measured amino acids. The exponential dependence of the probabilities on $\Delta G$ is consistent with an equilibrium distribution \cite{aminotherm}, although we caution that there are difficulties with this interpretation \cite{pizza}. Nevertheless, we take Eq. \ref{eq:empi} as our best approximation to the true equilibrium distribution. 
We furthermore assume that this function correctly predicts the equilibrium abundances of the heavier amino acids which have not yet been found in abiotic sources. This is consistent with the fact that it predicts abundances for these amino acids which would be too low to observe \cite{aminotherm}.

We take the distribution predicted from Eq. \ref{eq:empi} and \ref{eq:pro}, and compare it to the average non-equilibrium distribution, calculated numerically from Eq. \ref{eq:noneq}. We assume that amino acids are the most thermodynamically costly molecules that can be formed in the system. This ought to be the case if the system is physically confined to a small volume (e.g. a mineral pore), or the reactants are very diluted. Such a restriction on the available state space is needed because in the extreme non-equilibrium limit, all states become equally probable. This means that if more costly molecules can be formed than amino acids, the probabilities of forming any amino acids would go down relative to these more costly molecules. Nevertheless, the distribution of amino acids would become more uniform even without this restriction. In Sec. \ref{peptide} we relax this assumption on the maximum cost of molecules, as we look at the asymptotic behavior of amino acids polymerizing into arbitrarily long chains.

The average non-equilibrium distribution is plotted as a function of $\Delta G$ and compared to the equilibrium distribution in Figure \ref{fig:amino}, for various values of $\lambda$. Figure \ref{fig:tryrare} shows the probability of the rarest amino acid, tryptophan, as a function of $\lambda$. The concentrations of the rarest amino acids can be boosted by as many as 4 orders of magnitude in the non-equilibrium regime. Moreover, the roughly uniform distribution of amino acids employed in functional proteins is exactly what the average non-equilibrium distribution predicts in the extreme non-equilibrium regime (for values of $\lambda$ close to zero). Thus, far away from equilibrium, the distribution of amino acids moves closer to its biotic distribution, thereby greatly enhancing the chances of spontaneously assembling functional proteins \cite{typewriter,adami-information}.

\begin{figure}
\includegraphics[width=7cm,keepaspectratio]{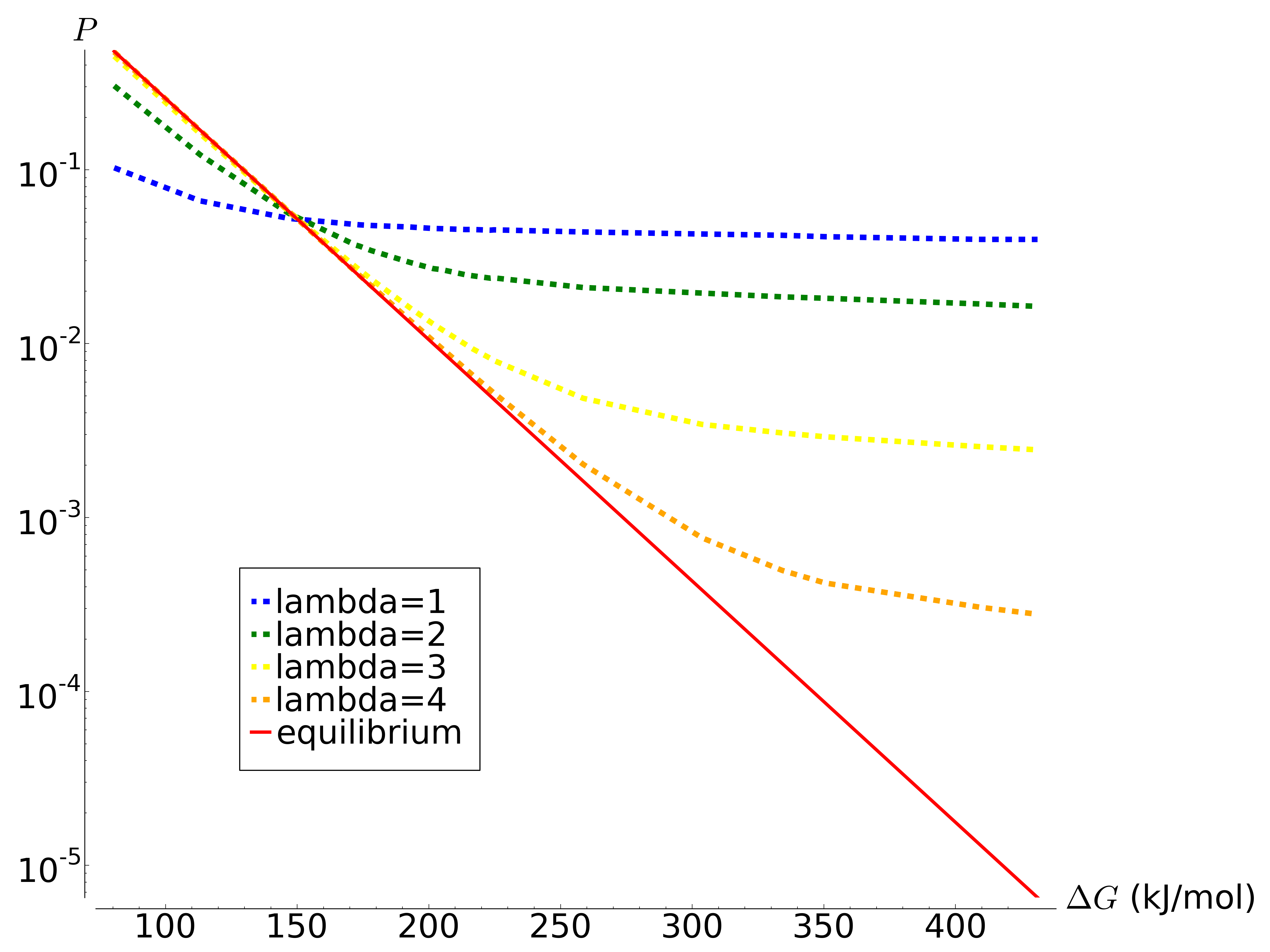}
\caption{The distribution of amino acids, arranged on the x-axis in order of increasing $\Delta G$. Each curve represents the average non-equilibrium distribution of amino acids given by Eq. \ref{eq:noneq}, at a different distance from equilibrium. Note that as the distance from equilibrium increases (i.e. $\lambda$ gets smaller), the distribution becomes flatter, with the probabilities of forming the rarest amino acids increasing by several orders of magnitude.}
\label{fig:amino}
\end{figure}

\begin{figure}
\includegraphics[width=7cm,keepaspectratio]{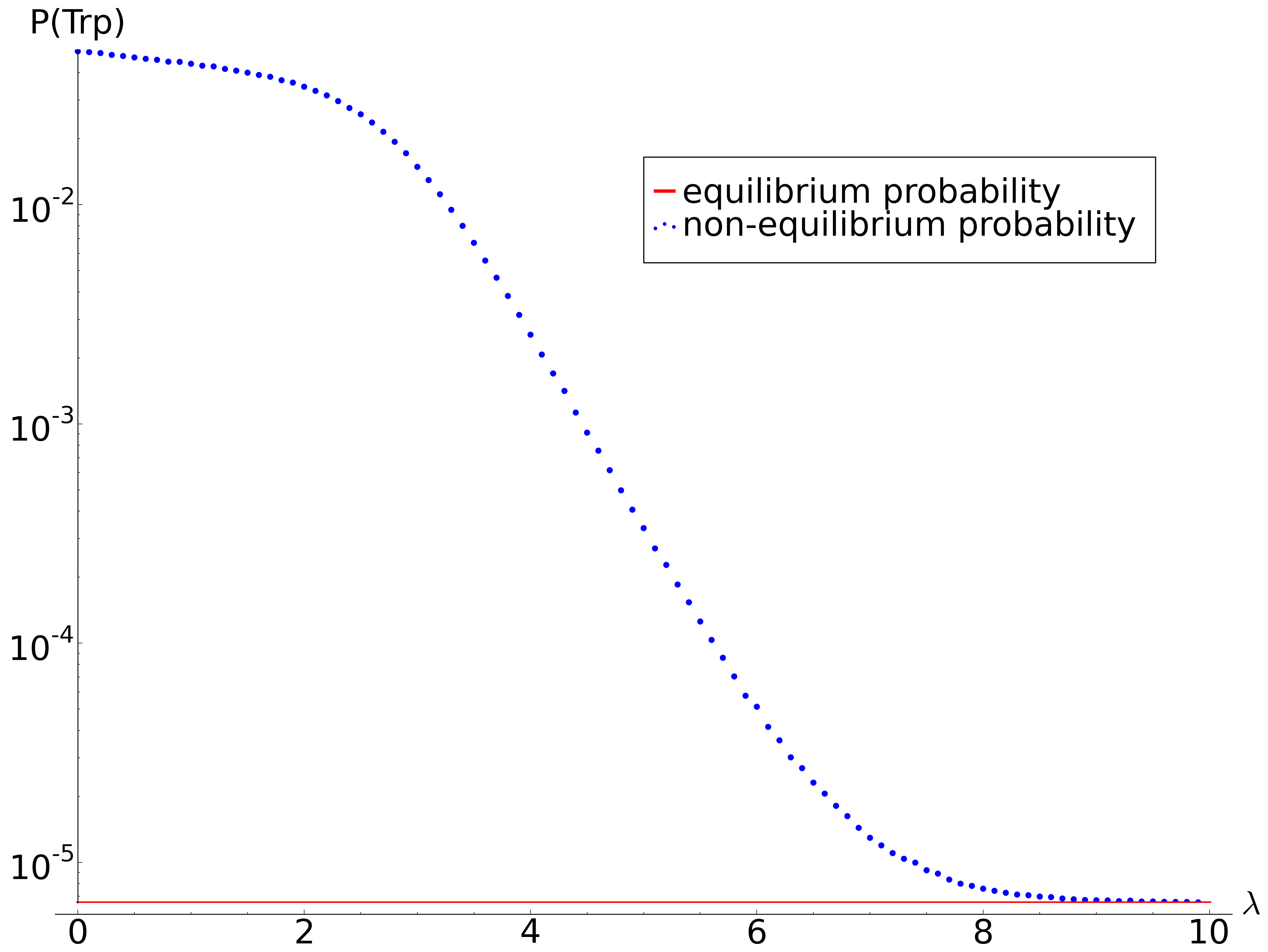}{\centering}
\caption{Tryptophan requires the largest free energy to form, and has not yet been found in an abiotic setting. Here we show how the concentration of tryptophan changes as one moves away from equilibrium, with the distance from equilibrium controlled by the parameter $\lambda$. We see that in the extreme non-equilibrium limit $\lambda \rightarrow 0$, the concentration of tryptophan can be increased up to a factor of $\sim 10^4$.}
\label{fig:tryrare}
\end{figure}


\section{Polymerization of Amino Acids} \label{peptide}

Amino acids may be linked with one another via the peptide bond to form long chains. These chains then fold into proteins, with a typical protein containing $\sim 500$ amino acids. However, $\Delta G$ for the peptide bond is on the order of several thousand kJ/mole \cite{peptide}, making the formation of long chains extremely improbable. It has been estimated that a solution containing $1M$ concentrations of each of the amino acids would require a volume $10^{50}$ times the size of the Earth to produce a single molecule of protein in equilibrium \cite{enzyme}. 

The thermodynamics of polymerization of amino acids were explored in \cite{peptide}, where, for simplicity, the chains were assumed to consist entirely of glycine. It was found that dimerization of two glycine molecules requires the greatest amount of free energy per bond ($\Delta G = 3.6$ kcal/mole), being about eight times more difficult to form than subsequent additions to the chain. The relative concentration $[GG]/[G]$ is predicted to be about $1/400$ in equilibrium, and each subsequent addition of a glycine to the peptide results in a decrease by a factor of $1/50$ \cite{peptide}. The probability of getting a chain of length $l \geq 2$ then follows a power-law

\begin{equation} \label{eq:pep}
 P_{\rm eq}(l) \propto \left(\frac{1}{50}\right)^{l-2}
\end{equation}

\noindent with the proportionality constant set by normalization. We examine the change in this distribution for non-equilibrium systems. To proceed, we identify each macrostate of a solution containing $N$ glycine molecules with a partition of the number $N$ into a sum of positive integers. For example, a solution containing $3$ glycine molecules could either be completely unbound, contain one dimer and one monomer, or one trimer.  For tractability, we consider only the extreme non-equilibrium limit $\lambda \rightarrow 0$ in this section, where all partitions of $N$ become equally likely. First, we examine the odds of the rarest state in equilibrium, where all $N$ glycine molecules become bound into a chain of length $l=N$. Then $P(l=N)=1/Q(N)$, where $Q(N)$ is the partition function. In number theory, the partition function $Q(N)$ counts the number of distinct ways that a positive integer $N$ can be decomposed into a sum of positive integers. We can estimate $P(l=N)$ using the Hardy-Ramanujan asymptotic expression for $Q(N)$ \cite{partitions}

\begin{equation} \label{eq:rama}
 P_{\rm neq}(l=N) \approx 4N \sqrt 3 * e^{-\pi \sqrt \frac{2N}{3}}.
\end{equation}

Clearly, the maximum probability of the rarest state is a decreasing function of $N$ in the $\lambda \rightarrow 0$ limit. Yet the odds of finding all $N$ particles bound into a single chain decrease much more rapidly in equilibrium, meaning that as the system gets larger, the factor by which non-equilibrium driving enhances probabilities of the rarest states grows without bound. This effect radically augments the chances of forming proteins in an abiotic setting. We display the ratio $P_{\rm neq}(l)/P_{\rm eq}(l)$ in Fig. \ref{fig:ratio}, using an exact expression for $P_{\rm neq}(l)$ obtained numerically.


        

\FloatBarrier

\begin{figure*}
\centering
\begin{minipage}[t]{.4\textwidth}
\includegraphics[width=7cm,keepaspectratio]{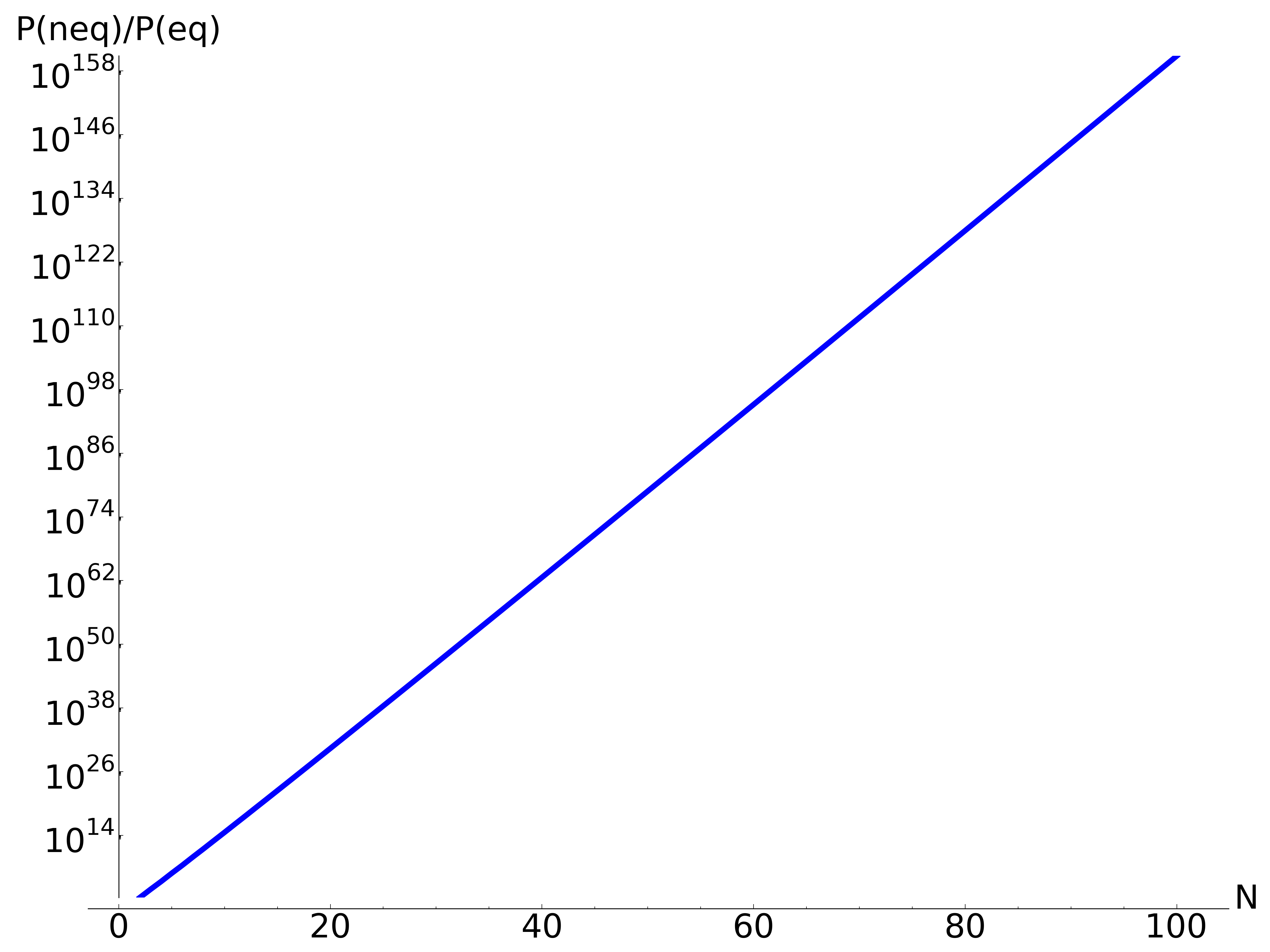}{\centering}
\caption{Glycine molecules may be linked together via a peptide bond to form chains. Due to the large amount of free energy required per bond, the concentrations of longer chains drop precipitously (Eq. \ref{eq:pep}). Here we consider a system of $N$ glycine molecules, and compute the ratio of finding all of them bound into a single long chain, in equilibrium and in the extreme non-equilibrium limit ($\lambda \rightarrow 0$). On the y-axis we display the non-equilibrium probability divided by the equilibrium probability. We see that as the number of molecules $N$ in the system grows, this ratio increases exponentially. This effect may help to explain how amino acids are spontaneously linked together to form proteins in an abiotic setting.}
\label{fig:ratio}
\end{minipage}\qquad
\begin{minipage}[t]{.4\textwidth}
\includegraphics[width=7cm,keepaspectratio]{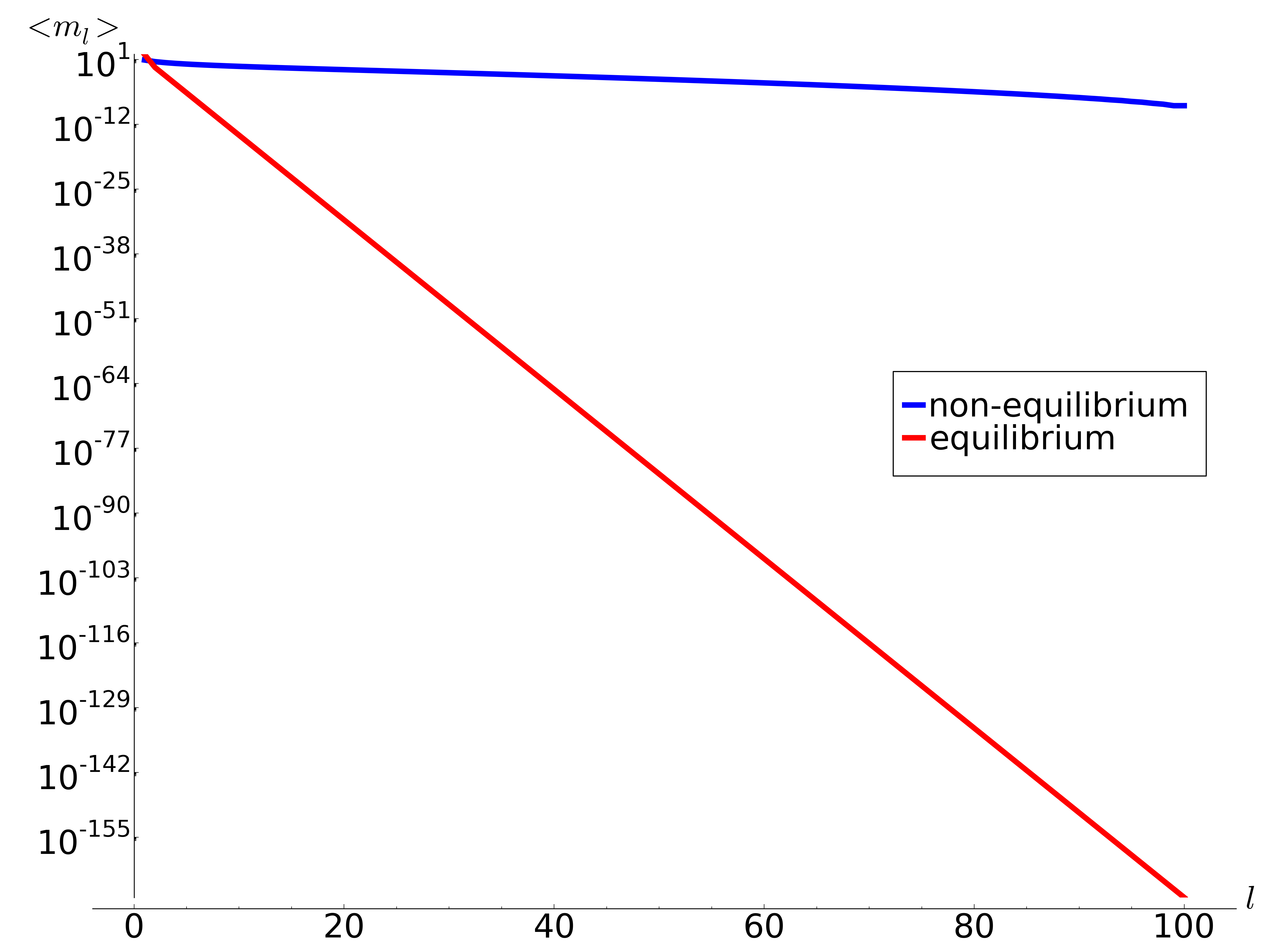}{\centering}
\caption{The expected number of chains of length $l$ in the extreme non-equilibrium limit is given by Eq. \ref{eq:chains}, while the equilibrium distribution is given by Eq. \ref{eq:pep}. These distributions are plotted for a system of size $N=100$, with the blue line representing the non-equilibrium case and the red line representing the equilibrium one. We see that the concentrations of the longest chains can be increased by hundreds of orders of magnitude out of equilibrium.}
\label{fig:chains}
\end{minipage}
\end{figure*}

Of interest is also the number of chains of length $l$, which we denote by $m_{l}$. When every partition is equally likely, the average number of chains of length $l$ is given by \cite{frag4,equipartitions}

\begin{equation} \label{eq:chains}
\langle m_{l} \rangle = \frac{1}{Q(N)}\sum\limits_{n=1}^{\rm int(N/l)} Q(N-nl).
\end{equation}

\noindent This distribution was previously studied in the context of a fragmentation process, e.g. where a nucleus is broken apart and each partition is equally likely \cite{equipartitions,frag1,frag2,frag3,frag4,frag5}. We calculate the set of $\langle m_l \rangle$ numerically for a system of size $N=100$ and compare to that predicted by Eq. \ref{eq:pep} in Fig. \ref{fig:chains}.


When $N$ is large and the chains aren't too long relative to $N$, Eq. \ref{eq:chains} is well approximated by \cite{frag4}

\begin{equation} \label{eq:asymp}
\langle m_{l} \rangle \approx \frac{1}{\mbox{exp} \left[\sqrt{\frac{\pi^2}{6N}} l \right]-1}
\end{equation}

\noindent which again will drop off much more slowly than the equilibrium distribution. This behavior means that in the extreme non-equilibrium limit, the chances of forming long peptide chains, and subsequently proteins, can be increased by hundreds of orders of magnitude. This shows one mechanism by which the spontaneous formation of proteins on the early Earth, which is all but excluded in equilibrium statistical mechanics, could become a viable possibility. This argument may thus bridge one of the more formidable barriers on the way to life's emergence.  

\section{Discussion}

We have demonstrated with two examples for which equilibrium thermodynamics seem to prohibit the pre-biotic synthesis of biologically important molecules, that under very modest assumptions, the concentrations of these molecules can be increased by many orders of magnitude when considering the average non-equilibrium distribution. This may well help to explain how environments on pre-biotic Earth might have produced the chemical precursors to life. A model-independent approach to assessing the odds of life's formation was also made in \cite{astrobio}, where the chances of emergence on other worlds was calculated from estimating parameters in a Drake-type equation. One of the parameters appearing in this equation is the abiogenesis probability $P_a$, which estimates the chances of life forming per unit time within a set of building blocks. An implication of our work is that this parameter ought to be increased on planets where these systems of building blocks are likely kept far from equilibrium, as for example on planets with rich weather phenomena, tectonic activity, or tidal interactions \cite{geodiseq}. The necessity for chemical disequilibrium on a planetary scale has been identified by several authors \cite{eric,geodiseq,drivetolife}, and the average non-equilibrium distribution $\langle \theta \rangle$ gives us a way of quantifying this effect as a function of $\lambda$.

Explaining the formation of heavy amino acids and peptides is, of course, far from completing the whole story. But we wish to emphasize that the average non-equilibrium distribution's increased odds for attaining otherwise rare states should be independent of the details of any particular reaction. Thus, the same effect is likely to play an important role in other situations where equilibrium thermodynamics create barriers to the emergence of life, e.g. the polymerization of nucleotides in RNA and DNA \cite{copoly}. It's also possible that the effect might be compounded, if for example a more favorable distribution of amino acids is input into another non-equilibrium system where they are assembled into peptides, and so on. Moreover, the biological relevance of this effect need not be limited to the origin of life. Indeed, it is possible that early metabolic processes drove intracellular molecular distributions even further from equilibrium, creating a feedback process whereby the state-space of useful molecules could be more effectively searched. A similar effect can be observed in kinetic proofreading, where energy is expended to drive reactions out of equilibrium and reduce the rate at which disadvantageous molecules are formed \cite{kinetic}. The degree by which the odds are increased depends on the value of $\lambda$, i.e. on how far from equilibrium the system has been driven.  

Altogether, our work raises the possibility that the formation of life does not require a particular environment that has been fine-tuned for life, but rather a set of environments which have been driven far enough away from equilibrium that obtaining favorable conditions becomes likely. Not only is life an inherently non-equilibrium phenomenon, but non-equilibrium driving may, in a general way, be the main catalyst for the emergence of life.

\section{Materials and Methods}

We usually do not know the exact configuration of systems with many degrees of freedom. Instead, we typically have access only to a few bulk characteristics of a system such as its pressure, volume, and temperature. 
Fortunately, statistical mechanics tells us that we do not need to describe the full microscopic state of the system in order to predict macroscopic characteristics, as those are understood as expectation values, or ensemble averages. Therefore, all we need to infer is the \textit{probability} of every state, $\rho_i$, $i=1, \dots N$. The problem, however, remains serious, as we have only a hand full of, say $M$, constraints, namely measured averages together with normalization of probability. So, we are still lacking $N-M$ equations to determine the $\rho_i$. Jaynes pointed out that equilibrium statistical mechanics assigns these probabilities by choosing that probability distribution with the largest entropy, $ S(\rho) \equiv - \sum_{i=1}^N \rho_i \ln (\rho_i)$, subject to the constraints imposed by the system's bulk properties \cite{Jaynes}. 

The maximization of entropy can be interpreted as choosing a model that makes use of only the information provided by the measured properties \cite{Jaynes, jaynes2003probability,gibbs2014elementary}.
This ensures that we do not ascribe to the system any information about its micro-states that we do not actually have. This powerful inference tool has since been applied to many other problems and is commonly known under the name of MaxEnt \cite{skilling2013maximum,kapur1989maximum}. In statistical physics, we find that under the constraint that only the average energy is known, the Boltzmann distribution is recovered: $\rho_i = \frac{1}{Z}(\beta) \exp{(-\beta E_i)}$. Boltzmann's constant $k_B$ scales inverse temperature, $\beta = 1/k_B T$, and $Z(\beta) = \sum_i \exp{(-\beta E_i)}$ is the partition function, ensuring normalization of the probability distribution.

It is much harder to infer the distribution, $\theta$, of a system that is away from thermodynamic equilibrium. The distribution can no longer be inferred straight from a MaxEnt argument, and information is lacking to make up for the missing equations.

One idea for circumventing this problem is to assign probabilities $P(\theta)$ to all distributions that might describe the system \cite{Crooks}. This means that our problem now becomes finding the \textit{distribution over distributions} that best describes the ensemble of non-equilibrium distributions, given the information we have about bulk properties of the system. Crooks proposed to use the distribution that maximizes the entropy, $S=-\int P(\theta)\ln P(\theta) d \theta$, subject to normalization, $\int P (\theta) d \theta = 1$, and subject to physically meaningful constraints. As such he used the average energy, which he writes as the expectation value $\langle \bar{E}(\theta) \rangle = \int P (\theta) \bar{E}(\theta) d \theta$, of the energies averaged over individual non-equilibrium distributions $\bar{E}(\theta) = \sum_i E_i \theta_i$, arguing that this does not add information beyond what is used to infer the equilibrium distribution. Additionally, he used the average entropy, $\langle S \rangle = \int P(\theta) S(\theta) d \theta$. This constraint introduces a measure of how far the system is from equilibrium, and the Lagrange multiplier used to enforce it parameterizes the deviation from the equilibrium distribution. The resulting distribution has the form \cite{Crooks} 

\begin{equation} \label{eq:thetapro}
P(\theta) = \frac{1}{\mathcal{Z}(\beta,\lambda)}\mbox{exp} \left[ -\lambda D (\theta \parallel \rho) \right] 
\end{equation}

\noindent where $\mathcal{Z}(\beta,\lambda)$ is a normalization constant.

At a fixed value of the Lagrange multiplier $\lambda$, a non-equilibrium distribution is more likely to occur, the closer it is to the equilibrium distribution in terms of the relative entropy (Eq. \ref{eq:thetapro}).  In the limit $\lambda \rightarrow \infty$, the equilibrium distribution attains a probability of one, and in the limit $\lambda \rightarrow 0$, all distributions become equally likely. 

The only assumption we are comfortable making about the conditions on early Earth is that the processes preceding life were {\em not} in thermodynamic equilibrium. 
We can not say anything about the details of the driving protocols, but it is reasonable to assume that conditions were inhomogeneous enough so that the specifics of the myriad of different non-equilibrium systems on pre-biotic Earth would average out. Let us therefore compare the probability of finding the building blocks of life as computed from the equilibrium distribution to that computed from the {\em average} non-equilibrium distribution. By integrating $\langle \theta \rangle = \int \theta P(\theta) d \theta$, we find 
\begin{equation}
\langle \theta \rangle = \frac{1}{\mathcal{Z}(\beta,\lambda)}\int \theta e^{ -\lambda D (\theta \parallel \rho)} d \theta ~ \label{thedist}
\end{equation}
(compare Eq. \ref{noneq}).

Numerical calculations were performed in SageMath. To calculate $\langle \theta \rangle$, we generated $20,000$ random distributions, then weighted them using Eq. \ref{eq:thetapro} and the given equilibrium distributions. We also added a sample of the equilibrium distribution to the set of random distributions, in order to correct for the possibility that no samples would be generated close enough to the equilibrium distribution to obtain appreciable weight, when $\lambda$ was high. Calculations for Fig.  \ref{fig:ratio} and \ref{fig:chains} were done exactly, using Sage's built in Partitions function.

\bibliographystyle{unsrt}
\bibliography{updatedreferences}

\begin{thebibliography}{10}

\bibitem{enzyme}
M~Dixon and EC~Webb.
\newblock Enzymes academic press.
\newblock {\em New York}, page 667, 1964.

\bibitem{typewriter}
Christoph Adami and Thomas LaBar.
\newblock From entropy to information: Biased typewriters and the origin of
  life.
\newblock {\em arXiv preprint arXiv:1506.06988}, 2015.

\bibitem{mineral}
Jean-Fran{\c{c}}ois Lambert.
\newblock Adsorption and polymerization of amino acids on mineral surfaces: a
  review.
\newblock {\em Origins of Life and Evolution of Biospheres}, 38(3):211--242,
  2008.

\bibitem{critparams}
HJ~Cleaves, AD~Aubrey, and JL~Bada.
\newblock An evaluation of the critical parameters for abiotic peptide
  synthesis in submarine hydrothermal systems.
\newblock {\em Origins of Life and Evolution of Biospheres}, 39(2):109--126,
  2009.

\bibitem{review}
Andr{\'e} Brack.
\newblock From interstellar amino acids to prebiotic catalytic peptides: a
  review.
\newblock {\em Chemistry \& biodiversity}, 4(4):665--679, 2007.

\bibitem{hydrotherm}
Everett~L. Shock and Mitchell~D. Schulte.
\newblock Organic synthesis during fluid mixing in hydrothermal systems.
\newblock {\em Journal of Geophysical Research: Planets},
  103(E12):28513--28527, 1998.

\bibitem{hydrorev}
William Martin, John Baross, Deborah Kelley, and Michael~J Russell.
\newblock Hydrothermal vents and the origin of life.
\newblock {\em Nature Reviews Microbiology}, 6(11):805--814, 2008.

\bibitem{alkavent}
Barry Herschy, Alexandra Whicher, Eloi Camprubi, Cameron Watson, Lewis
  Dartnell, John Ward, Julian R.~G. Evans, and Nick Lane.
\newblock An origin-of-life reactor to simulate alkaline hydrothermal vents.
\newblock {\em Journal of Molecular Evolution}, 79(5):213--227, 2014.

\bibitem{amend1998energetics}
JP~Amend and EL~Shock.
\newblock Energetics of amino acid synthesis in hydrothermal ecosystems.
\newblock {\em Science}, 281(5383):1659--1662, 1998.

\bibitem{Bada}
Jeffrey~L. Bada.
\newblock New insights into prebiotic chemistry from stanley miller{'}s spark
  discharge experiments.
\newblock {\em Chem. Soc. Rev.}, 42:2186--2196, 2013.

\bibitem{presynth}
Sherwood Chang.
\newblock Prebiotic synthesis in planetary environments.
\newblock In {\em The Chemistry of Life's Origins}, pages 259--299. Springer,
  1993.

\bibitem{eric}
Harold Morowitz and Eric Smith.
\newblock Energy flow and the organization of life.
\newblock {\em Complexity}, 13(1):51--59, 2007.

\bibitem{geodiseq}
LM~Barge, E~Branscomb, JR~Brucato, SSS Cardoso, JHE Cartwright, SO~Danielache,
  D~Galante, TP~Kee, Y~Miguel, S~Mojzsis, et~al.
\newblock Thermodynamics, disequilibrium, evolution: Far-from-equilibrium
  geological and chemical considerations for origin-of-life research.
\newblock {\em Origins of Life and Evolution of Biospheres}, pages 1--18, 2016.

\bibitem{gradient}
Christof~B Mast, Severin Schink, Ulrich Gerland, and Dieter Braun.
\newblock Escalation of polymerization in a thermal gradient.
\newblock {\em Proceedings of the National Academy of Sciences},
  110(20):8030--8035, 2013.

\bibitem{copoly}
David Andrieux and Pierre Gaspard.
\newblock Nonequilibrium generation of information in copolymerization
  processes.
\newblock {\em Proceedings of the National Academy of Sciences},
  105(28):9516--9521, 2008.

\bibitem{rob}
Robert~S. Shaw, Norman Packard, Matthias Schroter, and Harry~L. Swinney.
\newblock Geometry-induced asymmetric diffusion.
\newblock {\em Proceedings of the National Academy of Sciences},
  104(23):9580--9584, 2007.

\bibitem{Miller245}
Stanley~L. Miller and Harold~C. Urey.
\newblock Organic compound synthes on the primitive eart.
\newblock {\em Science}, 130(3370):245--251, 1959.

\bibitem{Crooks}
Gavin~E. Crooks.
\newblock Beyond boltzmann-gibbs statistics: Maximum entropy hyperensembles out
  of equilibrium.
\newblock {\em Phys. Rev. E}, 75:041119, Apr 2007.

\bibitem{drivetolife}
Michael~J Russell, Laura~M Barge, Rohit Bhartia, Dylan Bocanegra, Paul~J
  Bracher, Elbert Branscomb, Richard Kidd, Shawn McGlynn, David~H Meier,
  Wolfgang Nitschke, et~al.
\newblock The drive to life on wet and icy worlds.
\newblock {\em Astrobiology}, 14(4):308--343, 2014.

\bibitem{shaw1984dripping}
Robert Shaw.
\newblock {\em The dripping faucet as a model chaotic system}.
\newblock Aerial Press, 1984.

\bibitem{takara}
K~Takara, H-H Hasegawa, and DJ~Driebe.
\newblock Generalization of the second law for a transition between
  nonequilibrium states.
\newblock {\em Physics Letters A}, 375(2):88--92, 2010.

\bibitem{kullback1959}
S~Kullback.
\newblock Statistics and information theory.
\newblock {\em J. Wiley and Sons, New York}, 1959.

\bibitem{elements}
Thomas~M Cover and Joy~A Thomas.
\newblock {\em Elements of information theory}.
\newblock John Wiley \& Sons, 2012.

\bibitem{pizza}
Sandra Pizzarello, Yongsong Huang, and Megan Fuller.
\newblock The carbon isotopic distribution of murchison amino acids.
\newblock {\em Geochimica et Cosmochimica Acta}, 68(23):4963--4969, 2004.

\bibitem{McCollom}
Thomas~M. McCollom.
\newblock Miller-urey and beyond: What have we learned about prebiotic organic
  synthesis reactions in the past 60 years?
\newblock {\em Annual Review of Earth and Planetary Sciences}, 41(1):207--229,
  2013.

\bibitem{Adami}
Evan~D. Dorn, Kenneth~H. Nealson, and Christoph Adami.
\newblock Monomer abundance distribution patterns as a universal biosignature:
  Examples from terrestrial and digital life.
\newblock {\em Journal of Molecular Evolution}, 72(3):283--295, 2011.

\bibitem{adami-information}
Christoph Adami.
\newblock Information-theoretic considerations concerning the origin of life.
\newblock {\em Origins of Life and Evolution of Biospheres}, 45(3):309--317,
  2015.

\bibitem{aminotherm}
Paul~G Higgs and Ralph~E Pudritz.
\newblock A thermodynamic basis for prebiotic amino acid synthesis and the
  nature of the first genetic code.
\newblock {\em Astrobiology}, 9(5):483--490, 2009.

\bibitem{peptide}
R.~Bruce Martin.
\newblock Free energies and equilibria of peptide bond hydrolysis and
  formation.
\newblock {\em Biopolymers}, 45(5):351--353, 1998.

\bibitem{partitions}
George~E Andrews.
\newblock {\em The theory of partitions}.
\newblock Number~2. Cambridge university press, 1998.

\bibitem{frag4}
KC~Chase and AZ~Mekjian.
\newblock Nuclear fragmentation and its parallels.
\newblock {\em Physical Review C}, 49(4):2164, 1994.

\bibitem{equipartitions}
Joseph~R Iafrate, Steven~J Miller, and Frederick~W Strauch.
\newblock Equipartitions and a distribution for numbers: A statistical model
  for benford's law.
\newblock {\em Physical Review E}, 91(6):062138, 2015.

\bibitem{frag1}
Luciano~G Moretto and Gordon~J Wozniak.
\newblock The role of the compound nucleus in complex fragment emission at low
  and intermediate energies.
\newblock {\em Progress in Particle and Nuclear Physics}, 21:401--457, 1988.

\bibitem{frag2}
AZ~Mekjian.
\newblock Model of a fragmentation process and its power-law behavior.
\newblock {\em Physical review letters}, 64(18):2125, 1990.

\bibitem{frag3}
SJ~Lee and AZ~Mekjian.
\newblock Canonical studies of the cluster distribution, dynamical evolution,
  and critical temperature in nuclear multifragmentation processes.
\newblock {\em Physical Review C}, 45(3):1284, 1992.

\bibitem{frag5}
AS~Botvina, AD~Jackson, and IN~Mishustin.
\newblock Partitioning composite finite systems.
\newblock {\em Physical Review E}, 62(1):R64, 2000.

\bibitem{astrobio}
Caleb Scharf and Leroy Cronin.
\newblock Quantifying the origins of life on a planetary scale.
\newblock {\em Proceedings of the National Academy of Sciences},
  113(29):8127--8132, 2016.

\bibitem{kinetic}
J.~J. Hopfield.
\newblock Kinetic proofreading: A new mechanism for reducing errors in
  biosynthetic processes requiring high specificity.
\newblock {\em Proceedings of the National Academy of Sciences},
  71(10):4135--4139, 1974.

\bibitem{Jaynes}
E.~T. Jaynes.
\newblock Information theory and statistical mechanics.
\newblock {\em Phys. Rev.}, 106:620--630, May 1957.

\bibitem{jaynes2003probability}
Edwin~T Jaynes.
\newblock {\em Probability theory: The logic of science}.
\newblock Cambridge university press, 2003.

\bibitem{gibbs2014elementary}
J~Willard Gibbs.
\newblock {\em Elementary principles in statistical mechanics}.
\newblock Courier Corporation, 2014.

\bibitem{skilling2013maximum}
John Skilling.
\newblock {\em Maximum Entropy and Bayesian Methods: Cambridge, England, 1988},
  volume~36.
\newblock Springer Science \& Business Media, 2013.

\bibitem{kapur1989maximum}
Jagat~Narain Kapur.
\newblock {\em Maximum-entropy models in science and engineering}.
\newblock John Wiley \& Sons, 1989.

\end{thebibliography}


\end{document}